\documentclass[showpacs,amsmath,amssymb,amsfonts,preprint,prl,floatfix,superscriptaddress]{revtex4} 
\usepackage{graphicx}
\usepackage{bm}
\usepackage[german,english]{babel}

\begin{document}

\title{Trace of broken integrability in stationary correlation properties}

\author{Ioannis Brouzos}
\email{ioannis.brouzos@uni-ulm.de}
\affiliation{Institut f\"ur Quanteninformationsverarbeitung, Universit\"at Ulm, 89069 Ulm, Germany}

\author{Angela Foerster}
\email{angela@if.ufrgs.br}
\affiliation{Instituto de F\'{\i}sica da UFRGS, Av. Bento Gon\c{c}alves 9500, Porto Alegre, RS - Brazil}


\begin{abstract}
We show that the breaking of integrability in the fundamental one-dimensional model of bosons with contact interactions has consequences on the stationary correlation properties of the system. 
We calculate the energies and correlation functions of the integrable Lieb-Liniger case, comparing the exact Bethe-ansatz solution with a corresponding Jastrow ansatz.
Then we examine the non-integrable case of different interaction strengths between each pair of atoms by means of a variationally optimized Jastrow ansatz, proposed in analogy to the Laughlin ansatz.
We show that properties of the integrable state are more stable close to the Tonks-Girardeau regime than for weak interactions. 
All energies and correlation functions are given in terms of explicit analytical expressions enabled by the Jastrow ansatz. 
We finally compare the correlations of the integrable and non-integrable cases and show that apart from symmetry breaking the behavior changes dramatically, 
with additional and more pronounced maxima and minima interference peaks appearing.
\end{abstract}

\maketitle

The concept of integrability in physics originates in the earlier developments of the classical mechanics and goes back to Liouville ( see e.g. \cite{Arnold}).
One of the most familiar examples is the two-body Kepler problem, whereas the three body problem is not completely integrable. 
At the quantum level, the study of integrable systems has its origins in the work of Bethe in 1931 on the Heisenberg model \cite{Bethe}
and received a great impulse in the 1960s with the development of the exact solution of the Bose and Fermi gases with delta-function interaction, having prospered ever since \cite{Lieb}.
Although the precise definition of quantum integrability remains still an issue of discussion \cite{Caux}, 
some key ideas, such as the fundamental role played by the Yang-Baxter equation and the non-diffractive character of the Bethe ansatz wavefunction are well established in the field.
A significant aspect of integrable systems is that they can be found in different areas, such as statistical mechanics \cite{Baxter}, quantum field theory \cite{Faddeev}, 
condensed matter \cite{Essler}, nuclear physics \cite{Iachello}, string theory \cite{Zarembo} and more recently in cold atoms \cite{Eric}.

The developments in cooling and trapping atoms in optical lattices brought this subject to a new audience, when it became evident that integrable systems can be realized in the lab \cite{Batchelor, Kuhn, Guan, Batchelornew}.
Prominent examples include the Lieb-Liniger model \cite{Lieb}  for spinless bosons and the Yang-Gaudin model \cite{Yang, Gaudin} for two-component fermions, 
 or the McGuire impurity model in 1D Fermi gas recently realized experimentally \cite{McGuire, Jochim, Astra-Brouzos}.
In this scenario, the remarkable discovery that one-dimensional Bose gas of $^{87}\mathrm{Rb}$ do not thermalize after thousands of collisions \cite{Kinoshita} led to a natural question: what happens when integrability is broken? 
There has been a lot of activity in this direction related to the study of the quantum dynamics of integrable systems, 
in particular the behavior of the prototypical Lieb-Liniger  model after a quench \cite{Demler, Caux2} and a generalized Gibbs ensemble has been proposed \cite{Rigol}.  
The break of integrability has also found some implications in spin chains and transport properties \cite{Jung, Affleck}. 
Another possibility not much explored yet is how the  stationary many-body correlations are affected when the integrability is broken. 
Some differences are expected since Lamacraft constructed a wave-function that comprises a non-diffractive  and a diffractive part \cite{Lamacraft, Sutherland} to describe weak violations of integrability.
But to which extent the correlations are affected by the violation of integrability is still an open issue.

In this work, we break the integrability of the Lieb-Liniger model by setting different interaction strengths between each pair of bosons and show the consequences on the correlation properties. 
We start with the integrable case and compare the main approaches of correlated many-body wave functions: the Bethe ansatz,  which is the exact solution for this case,
and the Jastrow ansatz, based on the analytical exact two-body (pair) function, which is proven to be a very good approach getting almost exact if treated variationally.
The two ansatzes agree perfectly in the extreme cases of non- and infinitely strongly interacting limits, showing a small deviation for intermediate interactions.
We then set up a Jastrow ansatz for the non-integrable case of unequal interaction strengths and discuss its accuracy. 
We also explore the behavior of the variational parameter, the exponent of the pair-function in the variational Jastrow ansatz, 
which plays a similar role as the corresponding exponent in the celebrated Laughlin ansatz \cite{Laughlin}. 
A remarkable outcome of our Jastrow ansatz approach is that it allows to derive explicit analytical formulas for the energies and correlations, surprisingly not only for the integrable but also for the non-integrable case.
We examine also the stability of the properties of the integrable case, i.e., we analyze  qualitatively and quantitatively weather they persist as we go far from integrability, 
showing that close to the Tonks-Girardeau limit \cite{girardeau} the integrable behavior is more stable than for weak interactions.
We compare the correlation properties in terms of two-body correlation functions  showing effects like symmetry breaking and pronounced additional peaks for the non-integrable case,
which can be detected by state-of-the art experiments.

\section{Lieb-Liniger model, integrability and ansatzes} 

The model 1D-Hamiltonian
\begin{equation}
\label{ham}
H = - \frac{1}{2} \sum_{i=1}^N \frac{\partial^2}{\partial  x_i^2} + \sum_{i <j \leq N} g_{ij}  \delta(x_i-x_j)
\end{equation}
has an experimental realization with $N$ bosonic atoms of the same mass $M$ but in different hyperfine states (such that the interaction strengths between each pair $g_{ij}$ may differ) 
confined in a quasi-1D tube of oscillator length $a_{\perp}$.
The lengths are scaled by $L$ which is the size of the system for periodic boundary conditions $\psi(...,x_i=0,...)=\psi(...,x_i=1 [L],...)$ for all $i=1,N$ and the energies by $\hbar^2/ML^2$. 
In the standard experimental setup \cite{Jochim} the interaction strengths are controlled either by magnetic Feshbach or by confinement induced resonances \cite{olshanii} since $g=g_{1D}/(\hbar^2/ML)$ with  
$g_{1D}= \frac{2\hbar^2 a_{3D}}{M a^2_{\perp}} \left(1-\frac{|\zeta(1/2)| a_{3D}}{\sqrt{2} a_{\perp}}\right)^{-1}$ where $a_{3D}$ is the 3D s-wave scattering length.

The model with $g_{ij}=g$ (for all $i,j$) was introduced by Lieb and Liniger \cite{Lieb}  
and it is integrable in the Bethe ansatz sense. A discussion of its exact spectrum and thermodynamics can be found in \cite{Lieb, Korepin}.
We resume from their Bethe-ansatz solution the ground state wave function (in units of $1/L$) for the two-body case: 
\begin{equation}
\label{LL-2p}
 \Psi_{12}=c_{12} \cos \left[k_{12} \left( |x_1-x_2|-\frac{1}{2} \right) \right]
\end{equation}
where $k_{12} \in [0,\pi]$ (in units of $1/L$) as $g_{12}$ increases, according to the  condition $k_{12}=2\tan^{-1}(g_{12}/2k_{12})$ and 
$c_{12}=\left(\frac{2k_{12}}{k_{12}+\sin k_{12}}\right)$ is a normalization constant.
For more than two atoms the corresponding expression becomes more complex, and it does not provide an easy handling for further derivations, except for the thermodynamic limit. 
This is a basic motivation to introduce a simpler ansatz, using for its construction in terms of correlated pair functions the analytical solution of the two-body case. 
This type of functions were first introduced by Jastrow \cite{Jastrow} to solve problems with short-range interactions in nuclear physics and have also been employed recently in \cite{Brouzos1, Brouzos2}
as a correlated pair-function ansatz for 1D trapped bosons and fermions. 
Here we propose a new Jastrow-type ansatz, which in its general form reads:
\begin{equation}
\label{psi}
 \Psi_J^v = C\prod_{i <j }^{P} {\left( \Psi_{ij}
\right)}^v  .
\end{equation}
Above $\Psi_{ij}$ is the two-body exact solution (Eq.~\ref{LL-2p}) with the substitution $1\to i,2 \to j$,  $P$ is the number of pairs and $C$ a normalization constant, which can be calculated analytically (see Appendix).
We have included the exponent $v$ to make the ansatz variational, inspired by the corresponding parameter of the Laughlin wave function in the context of quantum Hall effect \cite{Laughlin}. 
For $v=1$ we have the non-variational Jastrow ansatz, since the same two-body boundary conditions hold for all pairs and define our $k_{ij}$ parameters,  
while for $v \in Re$ the conditions are changed to $k_{ij}=2\tan^{-1}(g_{ij}/2 v k_{ij})$.
In the integrable case all interaction strengths are equal leading to the same $k$, but in the non-integrable case we have a different $k_{ij}$ corresponding to a different $g_{ij}$ for each pair $ij$. 

In this work we will apply this ansatz to the simplest case where we can observe the effects of broken integrability, namely the 3-body problem, 
which is also from classical mechanics to Efimov states \cite{efimov} a still very active topic of research connected to integrable systems. 
In the case of the integrable Lieb-Liniger model the 3-body solution was discussed in \cite{Muga} and it contains $3!=6$ terms while our Jastrow ansatz only half of them.
This is already an important simplification especially for large $N$ where the Jastrow ansatz scales simply with the number of pairs $N(N-1)/2$. 
We consider the case where two atoms are identical with bosonic symmetry and interact with strength $g'$ (corresponding to $k'$ pair in our ansatz)
and the third or impurity atom interacts with the two majority atoms with strength $g$ (corresponding to $k$ pair in our ansatz). 
This means that we could choose, for instance, $g_{12}=g_{13}=g$ and $g_{23}= g'$ in the Hamiltonian (Eq.~\ref{ham}), 
implying that particles $2$ and $3$ ($1$) would be considered as majority (impurity) atoms for this choice (the integrable case corresponds to $g=g' \Rightarrow k=k'$). 
Let us also note that in the case of $g' \to \infty$ we have two majority fermionized bosons which for the local properties behave as free fermions.
So our ansatz can be easily expanded to the mixture of fermionic atoms, where the few-body case with an impurity has been  shown recently to capture already many-body behavior \cite{McGuire, Jochim, Astra-Brouzos,guan3}.

\section{Energy behaviour and accuracy of the ansatz}

\begin{figure}
\includegraphics[width=15 cm,height=15 cm]{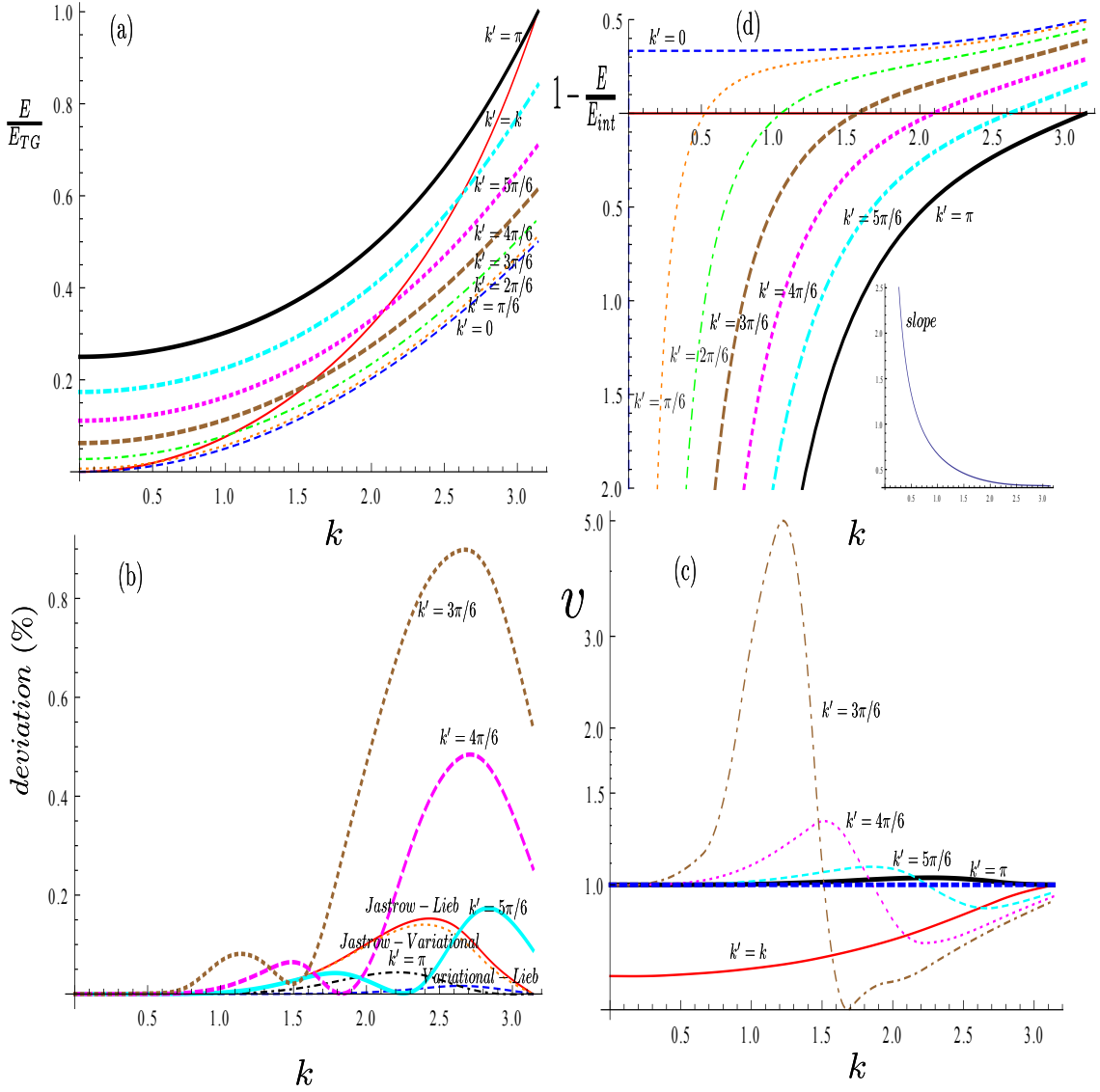}
\caption{(color online) (a)~Energy curves 
as a function of the Jastrow parameter $k$ 
for different values of $k'$, plotted from the analytical results (Eq.~\ref{energies}).
The integrable case $k=k'$ corresponds to the red smooth thin line.
(b)~Error for different approaches as a function of $k$: red smooth thin line - comparison of Jastrow and Bethe ansatz (exact solution) for the integrable case $k=k'$; 
 blue dashed thin line - comparison of variationally optimized Jastrow with Bethe ansatz for the integrable case; orange dotted thin line - comparison of  variational and non-variational Jastrow for the integrable case;
other lines - comparison of variational and non-variational Jastrow for non-integrable cases with different $k's$.
(c)~Behavior of the optimal value for the variational parameter $v$ 
as a function of $k$ for different $k's$. 
(d)~Stability of integrability: deviation of the energy from the integrable case ($1-E/E_{int}$) as a function of $k$ [labels as in (a), explained in the text].
Inset: slope of the deviation close to the integrable point as a function of $k$.}
\label{fig1}
\end{figure}

We first discuss the general behavior of the energy curves as a function of the interaction strengths for both the integrable and non integrable cases. 
Instead of the interaction strength we will use $k,k' \in[0;\pi]$, our basic Jastrow parameters. 
The transformation to the original variables $g,g'$ can be easily done due to the boundary conditions, as discussed in the previous section.
The non-variational Jastrow ansatz  ($v=1$) allows us to calculate the energy explicitly for both the integrable and non-integrable cases (see Appendix)

In Fig.~\ref{fig1}(a) the analytically calculated energy curves are shown as a function of the $k$ parameter of our ansatz for the integrable (red smooth thin line) and several non-integrable cases, corresponding to different $k'$ values. We are using $k'= \pi$ (black smooth thick line),
$k'= 5\pi/6$(cyan dashed-dotted thick line), $k'= 4\pi/6$ (magenta dotted thick line), 
$k'= 3\pi/6$ (brown dashed thick line),$k'= 2\pi/6$ (green dashed-dotted thin line), 
$k'= \pi/6$ (orange dotted thin line) and $k'=0$ (blue dashed thin line).
We observe the typical behavior of the Lieb-Liniger model with an increasing energy which saturates for the integrable case to $E_{TG}=4\pi^2$ [units $\hbar^2/ML^2$] of the Tonks-Girardeau limit at $g\to \infty$ ($k\to \pi$).
We normalize all curves to this limit value, which plays a significant role in 1D physics, since an infinitely strongly interacting system of bosons can be mapped to a non-interacting system of fermions \cite{girardeau}. 
We observe that all the curves for non-integrable cases cross the red curve at the integrable point where $k=k'$, while they lie above and below it for  $k<k'$ and $k>k'$, respectively.
This is expected since the total repulsive potential strength of the system is linearly dependent on $2g+g'$. 
Still, it is a first trace of the broken integrability which appears in the behavior of the energy.

At this level we perform a comparison between the different approaches to justify the validity of our Jastrow ansatz.
The starting point is the exact BA-solution for the energy in the integrable case, given by $E_B=\sum_i {\mathrm{k}}_i^2$, 
which obeys the boundary condition ${\mathrm{k}}_i=2[\pi-\tan^{-1}({\mathrm{k}}_i/g)-\tan^{-1}(2 {\mathrm{k}}_i/g)]$ \cite{Muga}. 
In general to find the energy for $N$ bosons one has to solve $N$ coupled transcendental equations ${\mathrm{k}}_m+\sum_{n=1}^N 2\tan^{-1}({\mathrm{k}}_m-{\mathrm{k}}_n/g)=2\pi(m-(N+1)/2)$ 
to define the ${\mathrm{k}}_m$ parameters of the Bethe ansatz (by symmetry arguments there is a reduction to $(N-1)/2$  $(N/2)$ parameters for $N$ odd (even) \cite{guan2}, as happens here).  
This is another difference from our Jastrow ansatz which contains a single $k$ parameter for the integrable case.

If we plot the results obtained by the variational Jastrow or Bethe ansatz on the scale of Fig.~\ref{fig1}(a), the curves would seem to overlap. 
This is a very impressive result, showing that the Jastrow ansatz, besides of having a very simple form and providing analytical explicit expressions for the energies of integrable and non-integrable cases,
provides a very good description for all cases, even when not treated variationally, at least to the order which is experimentally relevant.
To prove this statement we plot in Fig.~\ref{fig1}(b) the relative error of the Jastrow ansatz both in the integrable and the non-integrable cases. 
The red smooth thin line corresponds to a comparison of the Jastrow with the Bethe ansatz in the integrable case where the latter is the exact solution. 
We confirm also that the two approaches make exactly the same predictions in the extreme limits $g\to 0, g \to \infty$, 
because in both cases the Bethe ansatz exactly coincides with the Jastrow. This is another reason why to construct the Jastrow ansatz from this particular pair function. 
For a rather strong interaction $k \approx 4\pi/5$ (or equivalently $g \approx 15$) the deviation of the Jastrow ansatz energy from the exact solution $E/E_B-1$ takes its maximum value, 
[see red smooth thin curve (Fig.~\ref{fig1}(b))], which is nonetheless very small (lower than $0.2 \%$).
Still this is an indication that for the interaction range around this point the correlation properties deviate the most from the pair-like correlations of the Jastrow ansatz.

When we allow for a variational treatment of the Jastrow ansatz, by optimizing the parameter $v$, we observe that the agreement with the exact Bethe solution improves considerably, 
[blue dashed thin curve in Fig.~\ref{fig1}(b)], where the maximum deviation is lower than $0.01 \%$.
Therefore the non-variational Jastrow has a deviation of the same order of magnitude from the exact solution (red smooth thin curve) and from the variational Jastrow ansatz (orange dotted thin curve) in Fig.~\ref{fig1}(b).
This allows us to treat the variational Jastrow ansatz  as a very good approximation of the exact solution also in the non-integrable case.

Therefore lacking an exact solution in the latter case we compare the non-variational Jastrow ansatz with the variational one.
We observe that the deviation acquires its minimum values not only at very weak and very strong interactions, but also for an intermediate interaction strength 
(the additional minimum of the non-integrable cases: $k'= 3\pi/6$ (brown dotted thick line), 
$k'= 4\pi/6$ (magenta dashed thick line), $k'= 5\pi/6$ (cyan smooth thick line) and
$k'= \pi$ (black dashed-dotted thin line) in Fig.~\ref{fig1}(b)). 
This is another trace of the broken integrability scenario, which seems to approach a Jastrow type of correlations at three interaction strengths rather than two as in the integrable case.
One can further observe that the deviation of the non-variational Jastrow for these non-integrable cases is higher for larger $k$ but also for smaller $k'$.   

In Fig.~\ref{fig1}(c) we  plot also the behavior of the optimal parameter $v$ for the variational Jastrow ansatz as a function of $k$ for different values of $k'$. 
Here we observe that interestingly for the integrable case (red smooth thin curve) the optimal value of $v$ is always lower than $v=1$, which corresponds to the non-variational value, and also that it deviates more from 1 for low $k$.
For the non-integrable cases: $k'= 3\pi/6$ (brown dashed-dotted thin line), 
$k'= 4\pi/6$ (magenta dotted thin line), $k'= 5\pi/6$ (cyan dashed thin line) and
$k'= \pi$ (black smooth thick line) we see an oscillation of the variational parameter first to higher and then to lower than $v=1$ values. Also let us note that the deviation to higher values for lower $k'$ tends to go very high,
but this without improving  much the energy as we have seen in Fig.~\ref{fig1}(b) (corresponding  to a very smooth plateau for the variational optimization).

A further study for the energy behavior concerns the stability of the integrable case, i.e., the change of the energy as we  deviate from the integrable point $k=k'$ (by increasing or decreasing $k$). 
All the results above indicate already that there are different behaviors for different $k,k'$ values.
We observe in Fig.~\ref{fig1}(d) that the energy deviation ($1-E_{nonint}/E_{int}$) from the integrable $k=k'$ value (corresponding to the horizontal axis) is more abrupt for lower values of the interaction ($k'$). 
This is actually also possible to see from our Jastrow ansatz by calculating analytically the slope of the deviation at the vicinity of the integrable point as a function of the interaction strength $k$,
which is depicted in the inset of Fig.~\ref{fig1}(d). 
The explicit analytical expression is rather cumbersome, but from this figure we can see that the integrability is much more stable close to the TG-limit ($k\to \pi$) and very unstable close to zero interactions.
This stability mechanism of the integrability via strong repulsive interactions is one of the main results of our work.

\section{Correlation properties}

\subsection{General Remarks}

\begin{figure}
\includegraphics[width=15 cm,height=15 cm]{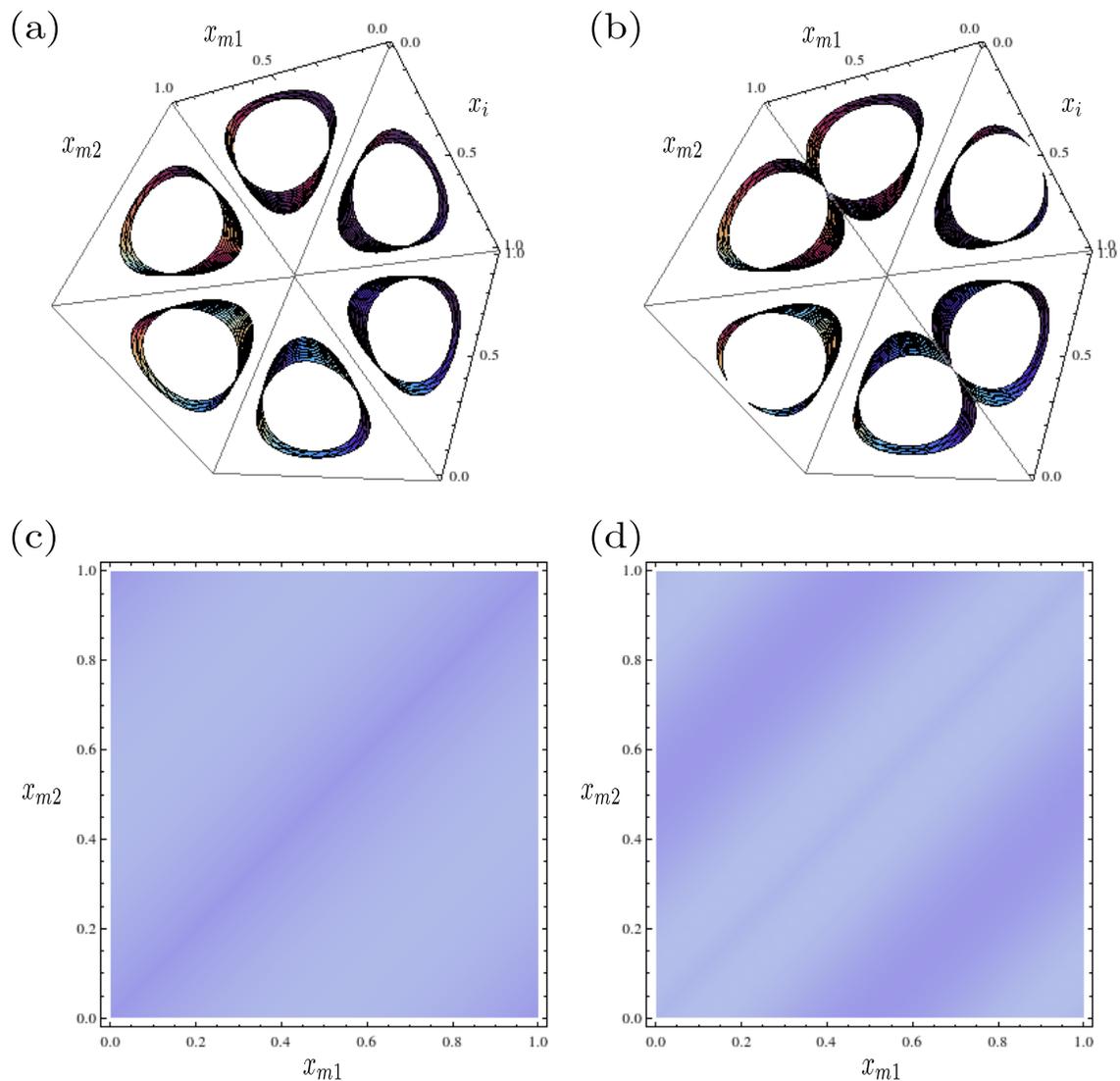}
\caption{(color online) Contour plot of the 3-body probability density at $|\psi|^2=0.007$,  shown from the perspective of a surface orthogonal to (1,1,1) vector for
(a) the integrable case $k=k'=5/2$; (b) the non-integrable case $k=5/2$ and $k'=3/2$, where the comparatively weaker interacting pair  of majority atoms is coming closer, breaking the symmetry.
Two-body density for the majority pair for (c) the integrable case $k=k'=\pi/3$ and (d) the non integrable case $k=5\pi/6$, $k'=\pi/3$, where pronounced additional minima appear. 
The light (dark) color indicates high (low) density.}
\label{fig2}
\end{figure}

The correlation properties are of great importance to understand the behavior of the system. 
Important findings concerning the correlations are summarized in Fig.~\ref{fig2} and will be further analyzed and elaborated in the next two figures and the following discussion.

In Fig.~\ref{fig2}(a) and (b) the full 3-body correlation function is shown projected as a contour plot for an integrable and a non-integrable case, respectively. 
The main consequence of violating integrability here is the breaking of symmetry, namely the fact that the weaker interacting pair is coming closer in Fig.~\ref{fig2}(b) (further discussion in Fig.~\ref{fig3}). 

In Fig.~\ref{fig2}(c) and (d) the two-body density $\rho(x_{m_1},x_{m_2})= \frac{\int |\psi|^2 dx_i}{\int |\psi|^2 dx_idx_{m_1}dx_{m_2}}$ 
of the majority pair of atoms is plotted for an integrable and a non-integrable case, respectively.
The main effect here is the additional very pronounced minima of this function on the wings around the diagonal. 
The diagonal ($x_{m_1}=x_{m_2}$) correlation hole is due to the repulsive contact interaction between this majority pair and is present in both cases.
However, when the impurity interacts strongly with the majority pair it produces a strong suppression of the density around the relative distance of $r=x_{m_1}-x_{m_2}=1/2$.
Due to the homogeneous space the two-body density is symmetric, so a complete characterization of the effects can be done just by cutting the off-diagonal section of  Fig.~\ref{fig2}(c) and (d)
which helps us to discuss in more detail the effects in  Fig.~\ref{fig4} and also in the Appendix Fig.~\ref{fig5}.


\subsection{Jacobi-coordinates picture}

\begin{figure}
\includegraphics[width=15 cm,height=15 cm]{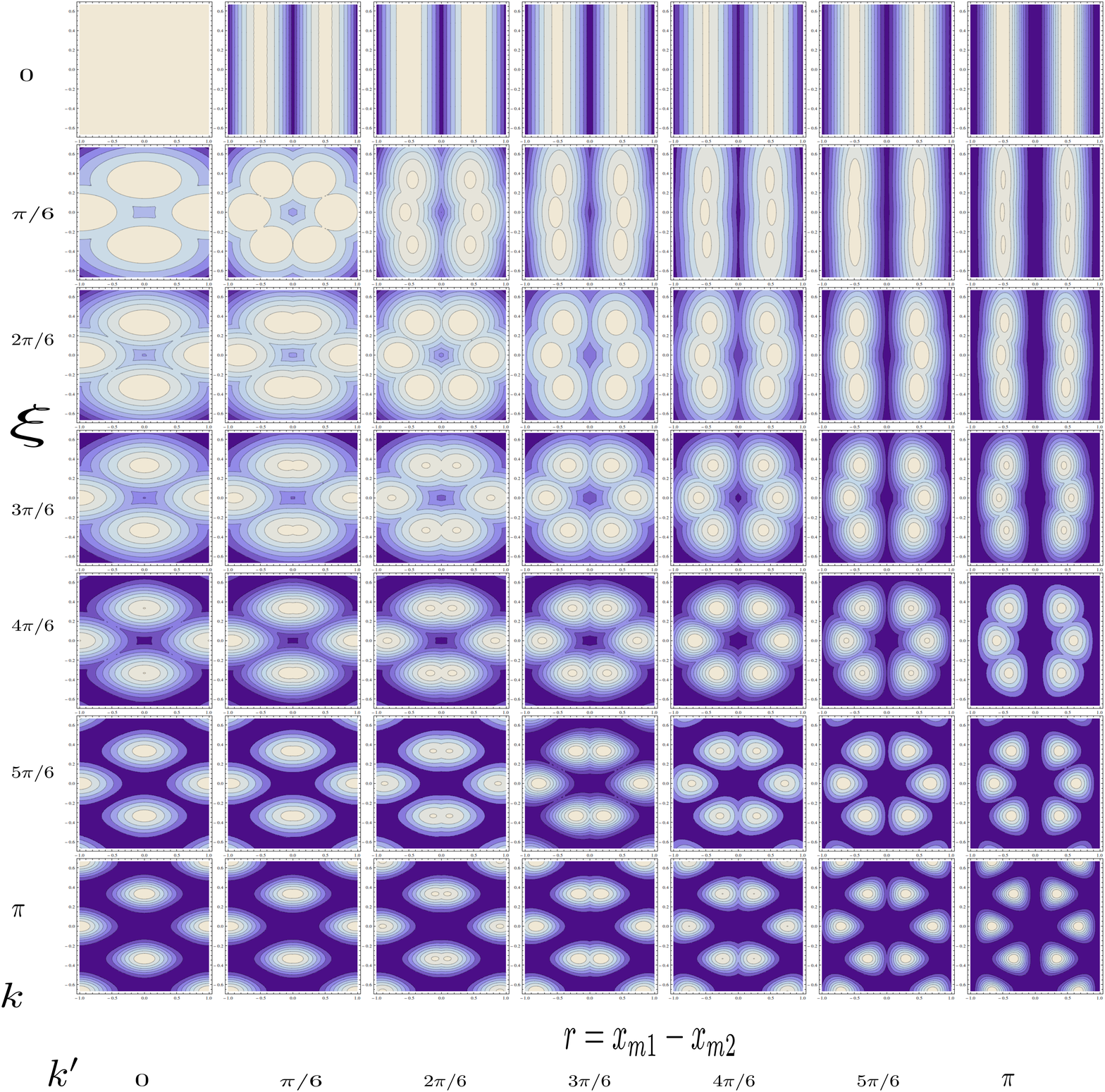}
\caption{(color online) Correlation functions in terms of Jacobi coordinates $\xi$ and $r$ for different values of $k$ and $k'$, depicted as a table. The integrable 
cases are presented on the diagonal of this table, with a symmetric pattern, in contrast to the non-integrable cases in the off-diagonals, where the symmetry  is broken.
The light (dark) color indicates high (low) values for the density. }
\label{fig3}
\end{figure}

Let us now discuss the consequences of violating integrability  in terms of correlations in  Jacobi coordinates, defined as: $\xi=\frac{2}{3}\left(x_i-\frac{x_{m_1}+x_{m_2}}{2}\right)$ and $r=x_{m_1}-x_{m_2}$. 
In  Fig.~\ref{fig3} we observe the effect of breaking the integrability in the off-diagonal plots of these block figures, which are all asymmetric compared to the diagonal ones, 
which in turn are completely symmetric and correspond to the integrable case $k=k'$. 
Notice also that the figures in the diagonal are very similar to the corresponding ones in the trapped case (see for instance \cite{Brouzos1}, \cite{Brendan}). 
This means, basically, that the trap does not produce a big impact in the central region of the correlation functions, where the contact repulsive potential is dominant.  
On general grounds, one may interpret each diagram as follows: the $\xi$ Jacobi coordinate represents the position of the impurity, depending on the position of the center of mass of the majority atoms.
Therefore the symmetry of the correlation function breaks when the interactions between the atoms are different, due to the change of preference in ordering the particles. 
In general the pairs experiencing stronger repulsions tend to avoid each other, also producing a rearrangement with the third particle.

Beyond these general remarks one can see in detail in Fig.~\ref{fig3} the following features:

i) On the uppermost row, where $k=0$, we have the two-body situation, since the  impurity does not interact with the majority atoms. 
As the interaction between the majority atoms increases, going  from left to right in this row, there is a depletion of the density (a  correlation hole) at the meeting point of the atoms ($r=0$). 
The profile along the vertical axis, representing the  position of the impurity with respect to the center of mass of the  majority atoms, is obviously homogeneous, due to the absence of interaction.

ii) This situation changes if we add the interaction with  the impurity; for instance, as we go from up to down in the first column,  
we notice that the impurity particle  tends to be at the center of  mass of the majority atoms if the latter are in a large distance, while it shifts to the edges if the majority atoms are close together. 
The  majority atoms tend to separate from each other if the interaction  with the impurity is strong and the impurity lies at their center of  mass. 
This is an interesting situation, since the majority atoms are  not directly coupled (in this column $k'=0$), but the indirect coupling via the impurity shifts the correlation profile, 
as well as that of the impurity itself with respect to their center of mass.

iii) As we move from left to right in all rows the main effect is  the appearance of the correlation hole in the middle of the vertical line at $r=0$ due to the increasing interaction between the majority  atoms. 
So the maxima which are still there in all left graphs and have  support at the line $r=0$ tend to split into two and separate completely  as $k'$ increases. 
Therefore one feature that differentiates most of the  plots of the bottom left triangle of this figure compared to the upper right one, i.e., the cases $k>k'$ and $k<k'$, respectively,
is the  separation of the majority pair, happening exactly at the integrable point. At the bottom left triangle we may have splitting  but not separation.

\subsection{Pair-correlation}

\begin{figure}
\includegraphics[width=15 cm,height=15 cm]{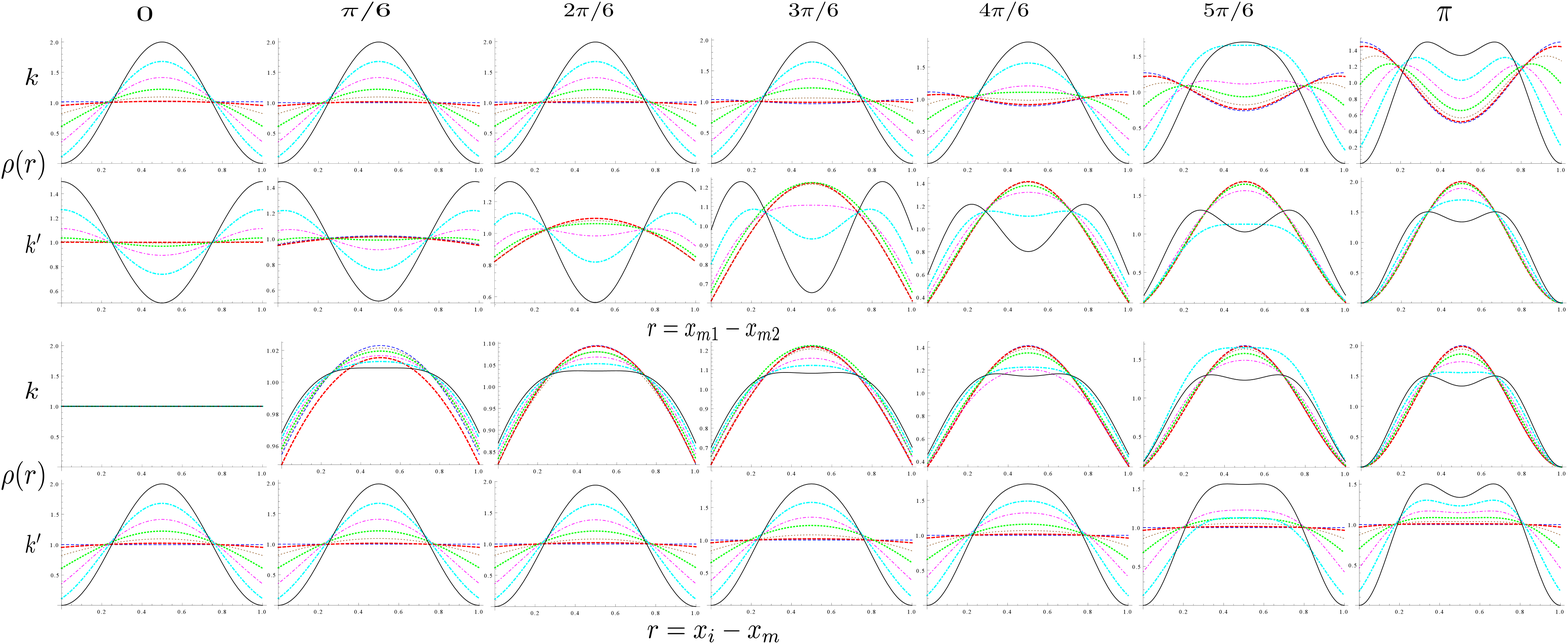}
\caption{(color online) The pair-correlation functions of the relative distance for the majority-majority pair (upper figures)  and the majority-impurity pair (lower figures).
As we move from left to right the $k$ or $k'$ parameter increases, according to the corresponding labels next to the leftmost figure of each row.
Within each figure the other $k$ or $k'$ parameter changes with the following color order of increasing $k$ or $k'$:  blue dashed thin line, red dashed thick line, green dotted thick line, brown dotted thin line, magenta dashed-dotted thin line, 
cyan dashed-dotted thick line, black smooth thin line. }
\label{fig4}
\end{figure}

We now consider the correlation function of the distance between two atoms,  which corresponds to the anti-diagonal cross section of Fig.~\ref{fig2} (c) and (d) (or also to integrating out the $\xi$ Jacobi coordinate). 
We provide in the Appendix the analytical formulas for both the integrable and non-integrable cases of the majority-majority  $\rho_{m_1m_2}(r=|x_{m_1}-x_{m_2}|)$ correlations 
as well as the impurity-majority ones $\rho_{im}(r=|x_i-x_m|)$. 

In Fig.~\ref{fig4} the obtained analytical expressions are ploted for representative values of $k,k'$, characterizing essentially the full two-body correlation behavior, whose main physical properties are 
the following:

i)  The Figure at the left uppermost corner represents the behavior of  the correlations of a two-body system. 
Here the interaction with  the impurity is turned off $(k=0)$ and the two majority particles interact with an increasing $k'$.
From $k'=0$ to $k'\to \pi$ we observe a smooth  transition from a completely homogeneous density profile, due to absence of interactions,
to a curved profile with minima at the meeting point of the strongly interacting pair at $r=0$ and $r=1$,
and maximum at $r=1/2$, which is the distance  that the atoms prefer to achieve in order to minimize their  interaction energy.

ii) Following the uppermost row of Fig.~\ref{fig4} we observe that the  two-body correlations of the majority atoms are not substantially  altered by the presence of weak interaction with the impurity atom.  
However, when the impurity-majority interaction exceeds $k>4\pi/6$, there is a  substantial change: the previously maximum at $r=1/2$ becomes now an  additional minimum.
When this minimum appears, the maxima shift to  $r=1/3$ (and $r=2/3$).
Then due to the strong  interaction with the impurity, which tends to take the position at a distance $r=1/2$ from the majority atoms, they in turn rearrange themselves in order to minimize the interaction energy with the impurity.
This behavior can also be noted in the rest of the figures and has been already demonstrated in Fig.~\ref{fig2}, 
while a more detailed discussion about the $k,k'$ values where this smooth transition happens can be found in the Appendix and the corresponding Fig.~\ref{fig5}.

iii) In each figure of the second uppermost row of Fig.~\ref{fig4} we can clearly see  the evolution and creation of additional minimum of the majority  pair correlation as the interaction with the impurity increases.
In the leftmost graph of this row the interaction between the majority  atoms is zero ($k'=0$) and therefore we observe only the impact of the  impurity. 
In this extreme case, where we start directly with  a homogeneous plateau, the minimum at $r=1/2$ appears immediately when  the interaction with the impurity is switched on. 
It is interesting to  observe the two uppermost left figures together, since they represent an opposite behavior: on the left corner, where we have two  atoms interacting with no effect from the impurity, 
the maximum arises at  $r=1/2$, while on the second row, where the impurity plays the role of an  indirect connection between two non-interacting atoms, exactly at $r=1/2$, we have a  minimum instead, 
and the $r=0, (r=1)$ positions are now preferred by  the majority atoms. 
It is also interesting to consider the rightmost  figure of this row, where we have the situation of infinite interaction or,
equivalently, fermionized  majority atoms ($k=\pi$), which is exactly the same as having two  non-interaction fermions, a relevant case for the experiments. 
Then the  fermionic nature of the correlations always pinned at zero when  the positions of the atoms coincide $(r=0)$, but still for quite strong  interaction strength with the impurity a minimum appears at $r=1/2$.

iv) The lower two rows in Fig.~\ref{fig4} represent the behavior of   impurity-majority correlations. 
Here we can point out the following effects: the impurity usually prefers to take a distance  $r=1/2$ from both of the majority atoms. 
The leftmost and bottommost  graphs represent exactly the same situation as before, i.e., when there is effectively only two atoms  interacting. 
This is always a  homogeneous case since there is no impurity-majority interaction  ($k=0$). 
We see that for quite large interaction strengths $k$ and $k'$, represented in the rightmost graphs of these rows,
the  impurity-majority correlation also acquires an additional (but not so  deep as before) minimum at $r=1/2$.
As expected, in the integrable  fermionized case (black smooth thin lines in the rightmost plots) the  profile  is identical for both majority-majority and majority-impurity  correlations. 
This is of course true in general for all integrable cases, while for all non-integrable cases we have an asymmetry between the two kinds of pair correlations.

\section{Concluding remarks and outlook}

We have shown how the breakdown of integrability in the fundamental one-dimensional model of bosons with contact interactions affects the stationary correlation properties of the three-body system in the repulsive regime. 
Besides of expected symmetry breaking effects, we found pronounced additional peaks in the correlations for the non-integrable case, which can be detected by state-of-the art experiments. 
In principle, a system of three particles with different interaction strengths can be experimentally realized by controlling inter- and intra species, thus providing an exciting opportunity to 
explore and test the nature of integrability with ultracold atoms.
We also presented a complete and detailed picture of the correlation functions for a large range of different interaction couplings (depicted in Figs. 3 and 4).  
In our analysis we introduced a new (variationally optimized) Jastrow ansatz, proposed in analogy to the Laughlin ansatz.
Remarkably, it allows to derive explicit analytical expressions for the energies and correlations, not only for the integrable but also for the non-integrable cases, which are very accurate. 
Our approach can be also extended to the attractive regime. In this case, however, a
a different scenario emerges, with the possibility of bound states. In principle, a corresponding Jastrow-type ansatz 
can be constructed, but then different two-body functions 
should be considered.
Our proposal can be employed to different systems, such as interacting fermions, bosons and mixtures composed of cold atoms with higher spin symmetry. 
In addition, it can also be adapted to handle a system of particles in a trap with different geometries, providing an alternative, efficient and much simpler way to existing methods. 
Finally our proposal can be directly used to discuss the case of a higher number of particles, which is presently an active research topic, specially in the discussion of the transition from few to many-body physics. 
In this context, our variationally optimized Jastrow ansatz could be particularly very useful to discuss the accuracy for higher N.

\acknowledgements
I.B. is thankful to L.Mathey and E.R.Ortega for helpful discussions.  
A.F. thanks M.T. Batchelor and X.W. Guan for inspiring discussions and CNPq
(Conselho Nacional de Desenvolvimento Cientifico e Tecnol\' ogico) for financial support.

\appendix
\section{Appendix}

The normalization constant for the 3-body Jastrow ansatz Eq.\ref{psi} is given by:  
    \begin{equation}
    C^2=  \left\{  
    \begin{array}{ll}
\frac{48 + 32 k^2 + 3 \cos a + 8 k^2 \cos k - 48 \cos 2 k - 3 \cos 3 k + 108 k \sin k}{256 k^2} ~~~~~~~~~~~~~~~~~~~~~~~~~~~~if~k=k' \\
       \frac {1}{(32 k' (k^3 - k k'^2)^2)}(4 k' ((k^2 - k'^2)^2 + k^4 \cos k') \sin^2 k +  8 k (k^2 - k'^2)^2 \sin k  (k' + \sin k') +  \\
   k (4 k k' (k^2 - k'^2)^2 + (4 k^5 - 10 k^3 k'^2 + 6 k k'^4 +  k'^2 (-3 k^2 + k'^2) \sin 2 k) \sin k' )) ~~~~\mathrm{else} \\
    \end{array}
    \right.
    \nonumber
    \end{equation}

The explicit analytical expression for the energy calculated by the non-variational Jastrow ansatz ($v=1$) of the integrable and non-integrable cases in terms of the ansatz parameters ($k,k'$) read:
\begin{equation}
\label{energies}
 E= \left\{
 \begin{array}{lll}
  3k^2  + 3k^2\left( 3 - 16\frac{ 3 + 2 k^2 + 2 k^2 \cos k - 3\cos 2k + 6 k \sin k}{(8 k^2- 1) \cos k + \cos 3k - 4 a \sin k}\right)^{-1}  ~~~~~if~k=k'\\
 2 k^2 +  k'^2 +  ~~~~~~~~~~~~~~~~~~~~~~~~~~~~~~~~~~~~~~~~~~~~~~~~~~~~~~ else \\
	\frac{3k^3 k'}{2} \frac{-2 k k' \cos k' \sin^2 k + (k^3 - k k'^2 + (k^2 + k'^2) \cos k \sin k ) \sin k'}
	{  ((k^2 - k'^2)^2 + k^4 \cos k') \sin^2 k + 8 k (k^2 - k'^2)^2 \sin k (k' + \sin k') +   
	k (4 k k' (k^2 - k'^2)^2 + (4 k^5 - 10 k^3 k'^2 +  6 k k'^4 + k'^2 (-3 {k}^2 + {k'}^2) \sin 2 k) \sin k'}
  \end{array}
\right.
 \end{equation}

The analytical formulas for  the integrable and non-integrable cases of the majority-majority  $\rho_{m_1m_2}(r=|x_{m_1}-x_{m_2}|)$ and impurity-majority $\rho_{im}(r=|x_i-x_m|)$ correlations read:

\begin{equation}
C^2\rho_{m_1m_2}(r)=\frac{\cos^2k'(r-1/2)}{16} [4k + 2kr \cos 2k(1-r)+ \sin 2k(1-r) + 2k(1-r)\cos 2kr + \sin 2kr + 8\sin k]
\end{equation}
which works for the integrable case by substituting $k'\to k$.


\begin{equation}
C^2\rho_{im}(r)=\left\{
\begin{array}{ll}
\frac{\cos^2(k(r-1/2)}{4 k k' (k^2-k'^2)} ~~~~~~~~~~~~~~~~~~~~~~~~~~~~~~~~~~~~~~~~~~~~~ k \ne k'\\ \cdot
[ (k^2 - k'^2 + k^2 \cos k'(1 - 2 r)) k' \sin k + k (k^2 - k'^2)(k'-\sin k')-k k'^2 \cos k(1-2r) \sin k']\\
\frac{\cos^2k(r-1/2)}{16} ~~~~~~~~~~~~~~~~~~~~~~~~~~~~~~~~~~~~~~~~~~~~~~k = k'\\ \cdot
[4k + 2kr \cos 2k(1-r)+ \sin 2k(1-r) + 2k(1-r)\cos 2kr + \sin 2kr + 8\sin k]
\end{array}
\right.
\end{equation}
As expected, we can observe that in the integrable case ($k=k'$) $\rho_{m_1m_2}(r)$ and $\rho_{im}(r)$ are identical, since there is no difference between majority and impurity atoms.


\begin{figure}
\includegraphics[width=6 cm,height=5 cm]{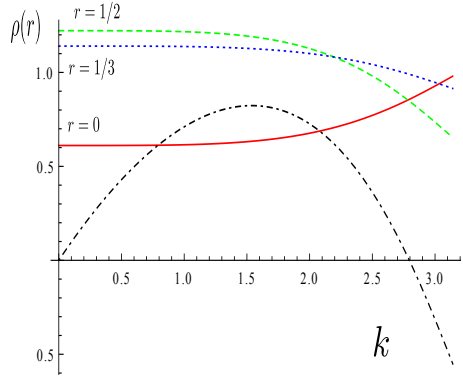}
\caption{(color online) Two-body correlations of the majority pair for different relative distances $r$ between the majority atoms as a function of $k$ for $k'=\pi/2$:
$r=1/2$ (green dashed line), $r=1/3$ (blue dotted line) and $r=0$ (red solid line).
The black dashed-dotted line corresponds to Eq.~\ref{corequation}, see discussion in the Appendix}
\label{fig5}
\end{figure}

A more detailed information about important features of the correlations can be obtained by looking at certain (crucial) relative distances.
In Fig.~\ref{fig5} we present the two-body correlations for different relative distances between the majority pair of atoms $r$.
More specifically, we plot $\rho_{m_1 m_2}(r=0),\rho_{m_1 m_2}(r=1/2), \rho_{m_1 m_2}(r=1/3)$ (red solid line, green dashed line, blue dotted line) as a function of $k$ for $k'=\pi/2$.
The crossing point of the green dashed line with the blue dotted line defines the point of smooth transition where the hole in $r=1/2$ appears [in Fig.~\ref{fig2}(d)], i.e.,  $\rho_{m_1 m_2}(1/2)=\rho_{m_1 m_2}(1/3)= \rho*$. 
For each $k$ there is a different value of $k'$ which we denote $k'*$ which corresponds to this point. 
We solve this equation explicitly and find $k'*$ and then we plot it in Fig.~\ref{fig5} (black dashed-dotted line):
 \begin{equation} 
\label{corequation}
k-k'* = k - 6 \arccos \left[\frac{\sqrt{6(2 k + k \cos k + 5 \sin k)}}{\sqrt{12 k + 4 k \cos (2 k/3) + 2 k \cos (4 k/3) + 
    3 \sin (2 k/3) + 24 \sin k + 3 \sin(4 k/3)}}\right]
\end{equation}
We plot with respect to the deviation from $k$, such that one can easily see that for intermediate values of $k$ and close to the TG limit one has to deviate more from integrability (increase $k'$) in order to see this transition.
An interesting finding here is that not only for very weak but also for intermediate to strong interactions (around $k=2.7$) the integrable case is very sensitive with respect to this transition.

\end{document}